\documentclass[letterpaper, 10 pt, conference]{IEEEtran}  

\IEEEoverridecommandlockouts                              

\renewcommand{\baselinestretch}{1}
\usepackage{graphics} 
\usepackage{epsfig} 
\usepackage{epic,eepic,epsfig}
\usepackage{epstopdf}
\usepackage{caption}
\usepackage{mathptmx} 
\usepackage{times} 
\usepackage{amsmath} 
\usepackage{amssymb}  
\usepackage{accents}
\usepackage{bm}

\usepackage{float}
\usepackage{pstricks}
\usepackage{subcaption}
\usepackage[normalem]{ulem}
\usepackage{hyperref}
\usepackage{multirow}

\usepackage{fancyhdr}

\def\lungo #1{\mathord{\buildrel{\lower3pt\hbox{$\scriptscriptstyle\frown$}}
\over #1 } }

\def\1D#1#2{{{\partial}\over{\partial #2}}#1}

\def\1d#1#2{{{d}\over{d #2}}#1}

\def\sign{\mathop{\rm sign}}

\newcommand{{\R}}{{\mathbb{R}}}
\newcommand{{\C}}{{\mathbb{C}}}

\def\sign{\mathop{\rm sign}}

\title{\LARGE{A Computationally Efficient Robust Model Predictive Control Framework for Ecological Adaptive Cruise Control Strategy of Electric Vehicles}
}

\author{Sheng Yu, Xiao Pan, Anastasis Georgiou, Boli Chen, Imad M. Jaimoukha and Simos A. Evangelou
\thanks{S. Yu, X. Pan, I. M. Jaimoukha and S. A. Evangelou are with the Dept. of Electrical and Electronic Engineering at Imperial College London, UK
        {\tt\small (sheng.yu17@ic.ac.uk, xiao.pan17@ic.ac.uk, i.jaimouka@ic.ac.uk, s.evangelou@ic.ac.uk)}}%
\thanks{A. Georgiou is with the Dept. of Mechanical Engineering at the University of Minnesota, USA
        {\tt\small (georg611@umn.edu)}}%
\thanks{B. Chen is with the Dept. of Electronic and Electrical Engineering at University College London, UK
        {\tt\small (boli.chen@ucl.ac.uk)}}%
}

\begin{document}

\thispagestyle{empty}
\setcounter{page}{0}
\begin{figure*}
\centering
\includegraphics[width=.9\textwidth]{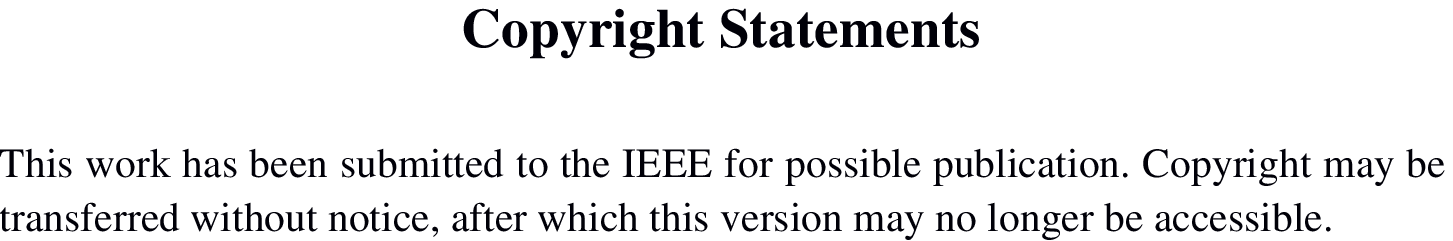}
\end{figure*}

\maketitle

\thispagestyle{fancy}
\chead{This work has been submitted to the IEEE for possible publication. Copyright may be transferred without notice, after which this version may no longer be accessible.}
\rhead{~\thepage~}
\renewcommand{\headrulewidth}{0pt}

\pagestyle{fancy}
\chead{This work has been submitted to the IEEE for possible publication. Copyright may be transferred without notice, after which this version may no longer be accessible.
}
\rhead{~\thepage~}
\renewcommand{\headrulewidth}{0pt}


\begin{abstract}
The recent advancement in vehicular networking technology provides novel solutions for designing intelligent and sustainable vehicle motion controllers. This work addresses a car-following task, where the feedback linearisation method is combined with a robust model predictive control (RMPC) scheme to safely, optimally and efficiently control a connected electric vehicle. In particular, the nonlinear dynamics are linearised through a feedback linearisation method to maintain an efficient computational speed and to guarantee global optimality. At the same time, the inevitable model mismatch is dealt with by the RMPC design. The control objective of the RMPC is to optimise the electric energy efficiency of the ego vehicle with consideration of a bounded model mismatch disturbance subject to satisfaction of physical and safety constraints. 
Numerical results first verify the validity and robustness through a comparison between the proposed RMPC and a nominal MPC. Further investigation into the performance of the proposed method reveals a higher energy efficiency and passenger comfort level as compared to a recently proposed benchmark method using the space-domain modelling approach.



\end{abstract}

\section{INTRODUCTION}
The development of intelligent and connected vehicles (ICV) aims to reduce potential risks of human-caused traffic accidents, and improve energy efficiencies and travel comfort levels~\cite{guanetti2018control}.
With real-time data collected by the sensors and cameras, the onboard controller of the ICV can optimally decide safety and energy-economic velocity trajectories. Existing studies of the ICV can be categorised into car-following~\cite{huang2018ecological,bae2019design} and intersection management scenarios~\cite{xu2021comparison,liu2020high}. 
As compared to the data-driven modelling technique utilising massive datasets and a learning-based approach for system identification and adaption~\cite{jia2021enhanced},
the present paper focuses on a model-based car-following problem, where the optimal trajectory is derived based on an accurate vehicle dynamics model and an inter-vehicular distance constraint to ensure safe driving conditions.

Most existing literature formulate the car-following problem as a nonlinear problem with consideration of the air-drag coefficient, rolling resistance and road gradient influences to minimise the model mismatch between the mathematics model and the real vehicle dynamics~\cite{7225178,GUO2022124732}.
In~\cite{7225178}, an ecological adaptive cruise controller aiming to improve both fuel economy and safety of hybrid electric vehicles whose dynamics are represented by nonlinear formulas dependent on air-drag resistance and road grade is developed. 
The work presented in \cite{GUO2022124732} combines car-following control and torque distribution control in a multi-objective nonlinear model predictive control (MPC) problem, where a nonlinear longitudinal dynamics model including engine rotational inertia is considered.
However, the main challenge for the potential application of nonlinear algorithms is the high computational burden, which hinders their real-world applications. Besides, depending on the initialisation point used within the solver sub-optimal solutions may be derived from the nonlinear optimisation problems.
To address the underlying issues, convex optimisation~\cite{sun2022tube} and linearisation techniques~\cite{8957499} are utilised for fast computational speed and unique global optimal solutions. \cite{sun2022tube} convexifies the inherently nonlinear MPC problem of heterogeneous vehicle platoons through coordination transformation and reasonable relaxations on non-convex constraints. A feedback linearisation method is adopted by~\cite{8957499} to turn the nonlinear vehicle dynamics into a linearised model as a part of its dual-layer distributed control scheme in an automated highway system.

In the authors' previous work, a convex framework for vehicle motion control has been developed using a space-domain modelling technique and state transformation~\cite{9838441,Pan_2022}. 
However, the space-domain method is not straightforward to be interfaced with other networked systems defined in the time-domain.
Therefore, this paper proposes an alternative time-domain control solution for car-following scenarios, while the convexity of the resulting optimisation problem is preserved by feedback linearisation of the nonlinear dynamics, including air-drag and rolling resistances. The optimisation is solved in a receding horizon manner by a robust model predictive control (RMPC), which also guarantees system robustness in the presence of modelling uncertainties introduced by unmodelled resistive forces. 
The contribution of the paper is threefold: 1) a feedback linearisation approach is proposed to transform the nonlinear electric-vehicle-participated car-following problem to a linear optimisation problem in the time-domain, 2) an RMPC method is adopted to address modelling mismatches that cannot be addressed by feedback linearisation, and 3) numerical examples reveal a higher energy efficiency and passenger comfort level of the proposed vehicle-following approach against a space-domain formulated benchmark~\cite{9838441} under the same simulation environment.

The rest of the paper begins with the modelling framework of the car-following problem, of which the formulation is convexfied through the feedback linearisation method in Section~\ref{sec:system description}. Section~\ref{sec:SDPR} introduces the time-domain formulated Semi-definite programming relaxation (SDPR) RMPC algorithm for the convex car-following problem. Simulation results are illustrated and discussed in Section~\ref{sec:simulation results}. Finally, conclusions are provided and a future work plan is suggested in Section~\ref{sec:conclusions}.  

\section{System and Control Problem Definition}
\label{sec:system description}
\subsection{Car-following Model}
\label{sec:car following model}
\begin{figure}[t!]
\centering
\includegraphics[width=\columnwidth]{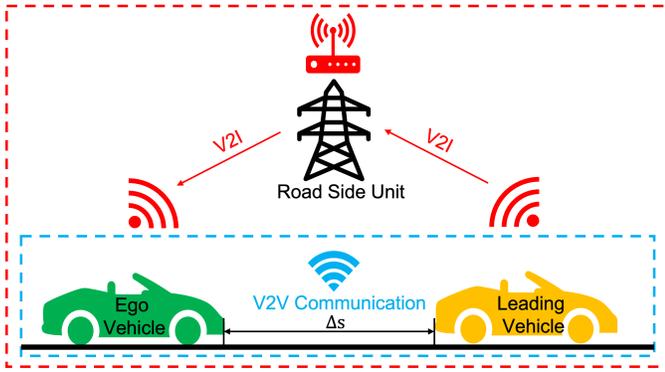}\\[-1ex]
\caption{Scheme of the car-following scenario with ego vehicle equipped with Vehicular ad hoc networks (VANET) communication. The ego vehicle can obtain the anticipated velocity trajectory of the leading vehicle either through direct V2V communication (the leading vehicle is connected) or relying on road side units to measure and transmit the leading vehicle data by V2I communication (the leading vehicle is not connected).}
\label{fig:V2V_time}
\end{figure}

This work focuses on the car-following problems with the two-way wireless vehicular ad-hoc network (VANET) communication technologies, as sketched in Fig.~\ref{fig:V2V_time}. 
In this context, the state information of the leading vehicle can be measured either through V2V or V2I depending on the intelligence and connectivity capability of the leading vehicle.
The ego vehicle, also denoted as the controlled vehicle, is a connected and automated vehicle, which can make control decisions based on the leading vehicle information.
In this work, it is assumed that the ego vehicle travels on the same lane as the leading vehicle and overtaking is not allowed. The information of the leading vehicle is assumed to be precisely measured and transmitted to the ego vehicle without communication errors and delays.
During the car-following paradigm as shown in Fig.~\ref{fig:V2V_time}, the longitudinal dynamics of the ego vehicle are considered and modelled as: 
\begin{multline}
    {v}(k+1)\!=\!v(k)\!+\!\left(\frac{F_w(k)}{m}\!-\!\frac{f_d(k)v^2(k)}{m}\right.\\
    \left.\!-\!gf_r(k)\cos{\theta}(k)-g\sin{\theta}(k)\right)\delta t,\label{eq:nonlinear_v_accurate}
\end{multline}
where the sampling index $k\in \mathbb{N}_{[0, \bar{k}]}$ with $\bar{k}\in\mathbb{N}_{>0}$ is the total number of samples. $\delta t\in\mathbb{R}_{>0}$ is the discretised sampling time interval. $m$ is the vehicle mass, $F_w(k)$ is the total force acting on wheels, $g$ is the Earth gravity, $f_r(k)$ and $f_d(k)$ are coefficients of rolling and air-drag resistances, respectively, influenced by real-time driving environments. $\theta(k)$ represents instantaneous road slope degrees. To quantitatively study and represent the modelling mismatch of the longitudinal dynamics shown above, \eqref{eq:nonlinear_v_accurate} can be rewritten as:  
\begin{multline}
    {v}(k+1)\!=\!v(k)\!+\!\left(\frac{F_w(k)}{m}\!-\!\frac{\tilde{f}_dv^2(k)}{m} \!-\!g\tilde{f}_r\!+\!w(k)\right)\delta t,\label{eq:nonlinear_v}
\end{multline}
where $\tilde{f}_d$ and $\tilde{f}_r$ are nominal coefficients and the nominal road slope is horizontal in this work, i.e. $\theta(k)=0$. $w(k)$ represents the modelling mismatch value between the nominal model and the actual model in \eqref{eq:nonlinear_v_accurate}, adding to the dynamics as the external bounded additive disturbance, satisfying $\underline{w}\leq w(k) \leq \overline{w}$ with its lower and upper bounds. The specification of the bounds of disturbance $w(k)$ in this work, which considers practical modelling mismatch, will be introduced later in Section~\ref{sec:simulation results}.

Moreover, the total force $F_{w}(k)$ of the ego vehicle with the electric motor is physically constrained by:
\begin{equation}
    F_{w,\min}\leq F_{w}(k)\leq F_{w,\max}, \label{eq:nonlinear_Fw_constraint_discrete}
\end{equation}
where $F_{w,\min}$ is the minimum force that can be applied on the wheel by the ego vehicle, which is consisted of the electrified powertrain output force as well as a mechanical braking force. $F_{w, \max}$ represents the maximum traction force executed by the electrified powertrain. For safety purposes, the speed limits are defined as:
\begin{equation}
    v_{\min}\leq v(k) \leq v_{\max},
    \label{eq:nonlinear_v_constraint}
\end{equation}
where $v_{\min}$ is the minimum allowed speed, and $v_{\max}$ is the maximum speed limit.

Consider $s_l(k)$ and $v_l(k)$, the position and velocity of the leading vehicle, respectively, such that:
\[
s_l(k+1) = s_l(k) + v_l(k)\delta t\,.
\]
The ego vehicle is expected to keep a safe headway distance $\Delta s(k)\!=\!s_l(k)-s(k)$ against the leading vehicle, which is defined as constraints with respect to the ego vehicle speed $v(k)$ and the minimum and maximum time gap, $\Delta t_{\min}$ and $\Delta t_{\max}$:
\begin{equation}
    \begin{aligned}
    \underbrace{s_{0}+v(k)\Delta t_{\min}}_{\Delta s_{\min}}\leq \Delta s(k) \leq \underbrace{s_{0}+v(k)\Delta t_{\max}}_{\Delta s_{\max}},\label{eq:nonlinear_delta_s_constraint}
    \end{aligned}
\end{equation}
where $s(k)$ are the positions of the leading and ego vehicles, respectively. $s_{0}$ is the standstill distance, $\Delta t_{\min}$ is the minimum time gap to avoid rear-end collision, and $\Delta t_{\max}$ is the maximum allowed time gap whose value is determined based on V2V communication range and traffic throughput efficiency~\cite{7659496,7374229}. For the sake of further discussion, the dynamics of $\Delta {s}(k)$ is also derived, as follows:
%
\begin{equation}
    \begin{aligned}
    \Delta {s}(k+1)=\Delta {s}(k)+(v_{l}(k)-v(k))\delta t\,.\label{eq:nonlinear_delta_s}
    \end{aligned}
\end{equation}


\subsection{Model Convexification}
\label{sec:model convexification}
The optimal control problem formulated above is non-convex due to the quadratic term ${\tilde{f}_dv(k)^2}$ in the air-drag resistance term of the system dynamics equation \eqref{eq:nonlinear_v}. Moreover, the rolling resistance only exists when the vehicle velocity is non-zero. Most existing research efforts address these non-convex issues by utilising the system state replacement technique in the space-domain~\cite{pan2022TCST,9151336} or simply considering a resistance loss vehicle model~\cite{9631299}. However, this work employs a feedback linearisation method to straightforwardly linearise the vehicle dynamics model~\eqref{eq:nonlinear_v} in the time-domain. The design of the linearisation feedback controller is given as follows:

\begin{equation}
    \begin{aligned}
    u_t(k)=F_w(k)-\tilde{f}_dv(k)^2-mg\tilde{f}_r \sign(v(k)).
     \label{eq:fl_new_u}
    \end{aligned}
\end{equation}


Due to the nonlinearity in the equation for $u_t(k)$, a conservative relaxation is made for the quadratic term $v(k)^2$ and the rolling resistance in \eqref{eq:fl_new_u}. The conservative constraints $u_{t,\min}(k)$, $u_{t,\max}(k)$ of $u_t(k)$ in \eqref{eq:fl_new_u} are given by:
\begin{equation}
    \begin{aligned}
    u_{t,\min}(k)\leq u(k) \leq u_{t,\max}(k), \label{eq:fl_constraint_relaxed}
    \end{aligned}
\end{equation}
with
\[
\begin{aligned}
u_{t,\min}&=F_{w,\min}-\tilde{f}_dv_{\min}^2\left(\geq F_{w,\min}-\tilde{f}_dv(k)^2-mg\tilde{f}_r\right),\\
u_{t,\max}&=F_{w,\max}-\tilde{f}_dv_{\max}^2-mg\tilde{f}_r\left(\leq F_{w,\max}-\tilde{f}_dv(k)^2\right).
\end{aligned}
\]

By substituting \eqref{eq:fl_new_u} into \eqref{eq:nonlinear_v}, the following linearised vehicle longitudinal dynamics in the time-domain can be obtained: 
\begin{equation}
    \begin{aligned}
    v(k+1)=v(k)+\left(\frac{u_t(k)}{m}+w(k)\right)\delta t. \label{eq:linear_delta_v}
    \end{aligned}
\end{equation}

\subsection{Model Predictive Control}
The nominal MPC control framework for the car-following problem is defined in this subsection, which will be instrumental for the introduction of the RMPC. 
To carry out the model predictive control design, the state-space equation of the linearised vehicle dynamics collecting \eqref{eq:nonlinear_delta_s} and \eqref{eq:linear_delta_v} in the time-domain within a receding horizon window with a prediction horizon length $N$ 
can be written as:
\begin{equation}
\begin{aligned}
 &x_t(k\!+\!1)=A_tx_t(k)\!+\!B_{tu}u_t(k)\!+\!B_{tc}C_t(k)\!+\!B_{tw}w(k),
    \label{eq:time_state_function_set}
\end{aligned}
\end{equation}
with
\[
\begin{aligned}
&x_t(k)\!=\!\begin{bmatrix}
    v(k)\\\Delta s (k)
    \end{bmatrix}, \quad A_t\!=\!\begin{bmatrix}
    1 \!\!&\!\! 0\\-\delta t \!\!&\!\! 1
    \end{bmatrix}, \quad B_{tu}\!=\! \begin{bmatrix}
    \frac{\delta t}{m}\\0
    \end{bmatrix}, \quad \\
    &B_{tw}\!=\!\begin{bmatrix}
    \delta t\\0 \end{bmatrix},\quad B_{tc}\!=\! \begin{bmatrix}
     0\\ \delta t
    \end{bmatrix}, \quad C_t(k)\!=\! \begin{bmatrix}
    v_{l}(k)
    \end{bmatrix},
\end{aligned}
\]
where $k\in \mathbb{N}_{[0, N-1]}$ with $N\in\mathbb{N}_{>0}$.
After incorporating the state bounds \eqref{eq:nonlinear_v_constraint} \eqref{eq:nonlinear_delta_s_constraint} as well as the control variable bound \eqref{eq:fl_constraint_relaxed}, the constraint function of the problem can be organised into a standard state-space form: 
\begin{equation}
\label{eq:time_constraint_function_set}
    \begin{aligned}
 &f_t(k)\!=\!C_{tf} x_t(k)\!+\!D_{tfu}u_t(k)\!+\!D_{tfw}w(k)\!,\\
 &f_t(k)=\begin{bmatrix}
    v(k)\\\Delta s (k)-\Delta t_{\max}v(k)\\\Delta s(k)-\Delta t_{\min}v(k)\\u(k)
    \end{bmatrix}, \; C_{tf}=\begin{bmatrix}
    1 & 0 \\ -\Delta t_{\max} & 1\\-\Delta t_{\min} & 1 \\0 & 0
    \end{bmatrix},\;\\
    &D_{tfu}= \begin{bmatrix} 0&0&0&1 \end{bmatrix}^{\top}, \;  D_{tfw}=\begin{bmatrix}0&0&0&0\end{bmatrix}^{\top},
   \end{aligned}
\end{equation}
which is constrained by lower and upper bounds, $\underline{f}_t(k)$ and $\overline{f}_t(k)$, respectively:
\begin{equation}
    \begin{aligned}
    \underline{f}_t(k)\leq f_t(k)\leq \overline{f}_t(k), \label{eq:time_constraint_function_set_inequality}
    \end{aligned}
\end{equation}
with
\[
\begin{aligned}
\underline{f}_t(k)&=[v_{\min},-\infty,s_{0}, u_{t,\min}]^{\top}, \\
\overline{f}_t(k)&=[v_{\max},s_{0},+\infty,u_{t,\max}]^{\top}.
\end{aligned}
\]
Note that the headway distance gap constraint~\eqref{eq:nonlinear_delta_s_constraint} is separated into two parts (i.e. the second and third terms in the bounds \eqref{eq:time_constraint_function_set_inequality}). This separation can be utilised to eliminate the impact of the disturbance when constructing the headway distance constraints, which is explained in detail as shown in Section~\ref{sec:SDPR}. 


By collecting the state-space equations of the state dynamics set \eqref{eq:time_state_function_set}, the constraint set \eqref{eq:time_constraint_function_set} and \eqref{eq:time_constraint_function_set_inequality}, the standard MPC problem with initial state $x_t(0)$ in the time-domain is formulated as: 
\begin{subequations}\label{eq:time_domain_problem}
\begin{align}
    &\hspace{12mm}\quad\min\limits_{u_t}\quad  V_t=\sum_{k=0}^{N} J_{t}(k)\,,  \label{eq:time_domain_problem_cost_function}\\
    &\textbf{subject to:}\nonumber\\ 
    &\hspace{6.5mm} x_t(k\!+\!1)\!=\!A_tx_t(k)\!+\!B_{tu}u_t(k)\!+\!B_{tc}C_t(k),
    \label{eq:time_domain_problem_state}\\
    & \hspace{6.5mm} \underline{f}_t(k)\leq f_t(k)\leq \overline{f}_t(k),\quad k=0,1,2,\ldots,N,
     \label{eq:time_domain_problem_constraint}  \\
    & \textbf{given: }\nonumber\\
    &\hspace{6.5mm} x_t(0)\!=\![v(0),\,\Delta s (0)]^{\top}, \label{eq:time_domain_problem_initial}
\end{align}
\end{subequations}
where $J_t(k)$ is a quadratic cost defined as:
%
\begin{equation}
\label{eq:time_cost_function_set}
\begin{aligned}
    &J_{t}(k)\!=\!(z_{t}(k)\!-\!\overline{z}_{t}(k))^{\!\!\!\top}\!\!Q_{t}(z_{t}(k)\!-\!\overline{z}_{t}(k)),\\
    &z_{t}(k)=C_{tz}x_t(k)+D_{tzu}u_t(k)+D_{tzw}w(k), \\
   & z_{t}(k)=\begin{bmatrix}
    v(k)\\ \Delta s (k)\\u_t(k)
    \end{bmatrix},
    \overline{z}_{t}(k)=\begin{bmatrix}
   \bar{v}\\\Delta s(0)\\0
    \end{bmatrix}, C_{tz}=\begin{bmatrix}
    1&0\\0&1\\0&0
    \end{bmatrix},\\
    & D_{tzu}=\begin{bmatrix}
    0,0,1
    \end{bmatrix}^{\top}, \quad D_{tzw}=\begin{bmatrix}0,0,0\end{bmatrix}^{\top}, 
\end{aligned}
\end{equation}
with the weighting matrix $Q_{t}\!=\!\text{diag}\{W_{1,t},W_{2,t},W_{3,t}\}\!\succeq\!0$. The $W_{2,t}$-term is a soft terminal constraint on the headway distance, aiming to keep the terminal distance gap equal to the initial gap, i.e. $\Delta s(0)=\Delta s(N)$, such that the activation of $W_{2,t}$ term only occurs when $k\!=\!N$ while $W_{2,t}\!=\!0$ for $k\in[0,N-1]$. 
Note here that the state dynamics \eqref{eq:time_domain_problem_state} does not include any disturbances, and the open-loop optimal control problem is written in a quadratic/convex form. Thus, the nominal MPC formulation can be solved by using a quadratic programming solver \cite{rawlings2017model}.






\section{Robust Model Predictive Controller}
\label{sec:SDPR}
    To tackle the disturbance $w(k)$ in \eqref{eq:time_domain_problem}, an RMPC scheme in the time-domain is proposed on the basis of on authors' previous work~\cite{9838441}, which formulated the optimisation control problem in the space-domain. In specific, the RMPC algorithm adopts the semi-definite programming relaxation to deal with bounded disturbances and reformulated the optimisation problem into linear matrix inequalities (LMIs). In addition to \cite{9838441}, the LMI-based RMPC approach has also been applied and verified in other industry areas \cite{Tahir13, 9304474,GEORGIOU202011974}. Comparisons between the new proposed RMPC in the time-domain and the previous proposed RMPC in the space-domain will be presented and investigated in the later Section~\ref{sec:simulation results}.
In order to formulate the control problem into an RMPC scheme, let us first define the following stack vectors:
\begin{equation}
\begin{aligned}
\label{eqn:RobustMPC_stack_signals}
    &\mathbf{x_t}\!=\!\begin{bmatrix} x_t(0)\\x_t(1)\\ \vdots \\x_t(N) \end{bmatrix}\!\!\!,\quad
    \mathbf{f_t}\!=\!\begin{bmatrix} f_t(0)\\f_t(1)\\ \vdots \\f_t(N) \end{bmatrix}\!\!\!,\quad
    \mathbf{z_t}\!=\!\begin{bmatrix} z_t(0)\\z_t(1)\\ \vdots \\z_t(N) \end{bmatrix}\!\!\!,\\
    &\mathbf{w}\!=\!\begin{bmatrix} w(0)\\w(1)\\ \vdots \\w(N\!-\!1)\end{bmatrix}\!\!\!,
     \mathbf{u_t}\!=\!\begin{bmatrix} u_t(0)\\u_t(1)\\ \vdots \\u_t(N-1) \end{bmatrix}\!\!\!,
    \mathbf{C_t}\!=\!\begin{bmatrix} C_t(0)\\C_t(1)\\ \vdots \\C_t(N-1) \end{bmatrix}\!\!\!.
\end{aligned}
\end{equation}

Analogous to our previous RMPC scheme in~\cite{9838441}, the convex optimisation problem in the time-domain is firstly formulated in a stacked form over prediction horizon $N$ as:\\[-2.6ex]
\begin{subequations}\label{eq:mpc_time_stack_problem}
\begin{align}
     \min\limits_{\mathbf{u_t}}\quad& \mathbf{J_{t}}\,,\\
    \textbf{s.t. } & \mathbf{x_t}=\widetilde{A}_t x_t(0)+\widetilde{B}_t \mathbf{u_t} +\widetilde{B}_{tc} \mathbf{C_t}+\widetilde{B}_{tw} \mathbf{w}\,,\\
    &\mathbf{\underline{f}_t}\leq \mathbf{f_t}\leq \mathbf{\overline{f}_t}\,,\label{eq:mpc_time_stack_constraint}\\
    \textbf{given: }& x_t(0)\!=\![v(0),\,\Delta s (0)]^{\top},
\end{align}
\end{subequations}
where $\widetilde{A}_t$, $\widetilde{B}_t$, $\widetilde{B_{tc}}$, $\widetilde{B}_{tw}$ are stacked coefficient matrices and are readily obtained from iterating the dynamics in \eqref{eq:time_state_function_set}. Moreover,  $\mathbf{\underline{f}_t},\,\mathbf{\overline{f}_t}\!\in\!\mathbb{R}^{4(N+1)\times 1}$ are the stacked vectors of upper and lower bounds $\underline{f}_t(k),\,\overline{f}_t(k)$, respectively, and
\begin{multline}
    \mathbf{J_{t}}\!=\!(\widetilde{C}_{tz}x_t(0)\!+\!\widetilde{D}_{tzu}\mathbf{u_t}\!+\!\widetilde{D}_{tzc}\mathbf{C_t}\!+\!\widetilde{D}_{tzw}\mathbf{w}\!-\!\mathbf{\overline{z}_t})^{\top}\mathbf{Q_t}\\
  (\widetilde{C}_{tz}x_t(0)\!+\!\widetilde{D}_{tzu}\mathbf{u_t}\!+\!\widetilde{D}_{tzc}\mathbf{C_t}\!+\!\widetilde{D}_{tzw}\mathbf{w}\!-\!\mathbf{\overline{z}_t})\,,
\end{multline}
where $\widetilde{C}_{tz}$, $\widetilde{D}_{tzu}$, $\widetilde{D}_{tzw}$ are stacked coefficient matrices after iterating the $J_t(k)$ equation in \eqref{eq:time_cost_function_set}, $\mathbf{\overline{z}_t}\in\mathbb{R}^{3(N+1)\times 1}$ is the stacked vector of $\overline{z}(k)$, and $\mathbf{Q_t}$ is the stacked matrix of the weighting matrix $Q_t$.

To solve the above optimisation problem, the computationally efficient and verified SDPR method is utilised by introducing a new auxiliary variable, $\overline{\gamma}_t$, which provides an upper bound of stacked cost functions $\mathbf{J_t}$: 
\begin{equation}
    \begin{aligned}
    \mathbf{J_t}- \overline{\gamma}_t\leq 0. \label{eq:RMPC_gamma}
    \end{aligned}
\end{equation}

After applying SDPR to the left-hand side (LHS) of the inequality \eqref{eq:RMPC_gamma}, we obtain the identity:
\begin{equation}
    \begin{aligned}
    LHS_{\eqref{eq:RMPC_gamma}}\!=\!-(\mathbf{w}-\underline{\mathbf{w}})^{\top}D_t(\mathbf{\overline{w}}-\mathbf{w})\!-\!\left[\mathbf{w}^{\top}\,\,1\right]L(\mathbf{u_t},D_t,
    \overline{\gamma}_t)\begin{bmatrix}\mathbf{w}\\1\end{bmatrix}\!, \label{eq:RMPC_gamma_LHS}
    \end{aligned}
\end{equation}
where $D_t\succeq 0$ with $D_t \in \mathbb{R}^ {N \times N}$ is a positive semi-definite diagonal matrix, and $L(\mathbf{u_t},D_t,\overline{\gamma}_t)$ is a matrix dependent on $\mathbf{u_t}$, $D_t$, and $\overline{\gamma}_t$:
\begin{equation*}
\begin{aligned}
&L(\mathbf{u_t},D_t,\overline{\gamma}_t)=\\
    &\begin{bmatrix}
    -\widetilde{D}_{tzw}^{\top}\!\mathbf{Q_t}\widetilde{D}_{tzw}\!+\!
    D_t & -D_t(\underline{\mathbf{w}}+\overline{\mathbf{w}})/{2}-bd_t \\
    * & \hspace{-7mm} \underline{\mathbf{w}}^{\!\!\!\top}\!D_t\overline{\mathbf{w}}\!-\!cd_t\!-\!\mathbf{u_t}^{\!\!\!\top}\!\widetilde{D}_{tzu}^{\top}\!\!\mathbf{Q_t}\widetilde{D}_{tzu}\mathbf{u_t}\!+\!\overline{\gamma}_t\end{bmatrix},\\
\end{aligned}
\label{eq:L1}
\end{equation*}    
with $*$ denotes the symmetry element of the corresponding matrix and   
\begin{equation*}
\begin{aligned}      
&bd_t\!=\!\widetilde{D}_{tzw}^{\top}\mathbf{Q_t}\widetilde{C}_{tz}x_{t}(0)\!+\!\widetilde{D}_{tzw}^{\top}\mathbf{Q_t}\widetilde{D}_{tzu}\mathbf{u_t}\\
&\quad\!+\!\widetilde{D}_{tzw}^{\top}\mathbf{Q_t}\widetilde{D}_{tzc}\mathbf{C_t}\!-\!\widetilde{D}_{tzw}^{\top}\mathbf{Q_t}\mathbf{\overline{z}_{t}},\\
&cd_t=x_{t}(0)^{\top}\widetilde{C}_{tz}^{\top}\mathbf{Q_t}\widetilde{C}_{tz}x_{t}(0)\!+\!x_{t}(0)^{\top}\widetilde{C}_{tz}^{\top}\mathbf{Q_t}\widetilde{D}_{tzu}\mathbf{u_t}\\
&\quad \!+\!x_{t}(0)^{\top}\widetilde{C}_{tz}^{\top}\mathbf{Q_t}\widetilde{D}_{tzc}\mathbf{C_{t}}\!+\!\mathbf{u_t}^{\top}\widetilde{D}_{tzu}^{\top}\mathbf{Q_t}\widetilde{C}_{tz}x_{t}(0)\\
&\quad\!+\!\mathbf{u_t}^{\top}\widetilde{D}_{tzu}^{\top}\mathbf{Q_t}\widetilde{D}_{tzc}\mathbf{C_{t}}\!+\!\mathbf{C_t}^{\top}\widetilde{D}_{tzc}^{\top}\mathbf{Q_t}\widetilde{C}_{tz}x_{t}(0) \\
&\quad\!+\!\mathbf{C_t}^{\top}\widetilde{D}_{tzc}^{\top}\mathbf{Q_t}\widetilde{D}_{tzu}\mathbf{u_{t}}\!+\!\mathbf{C_t}^{\top}\widetilde{D}_{tzc}^{\top}\mathbf{Q_t}\widetilde{D}_{tzc}\mathbf{C_{t}} \\
 &\quad\!-\!x_{t}(0)^{\!\!\top}\widetilde{C}_{tz}^{\top}\mathbf{Q_t}\mathbf{\overline{z}_t}\!-\!\mathbf{u_t}^{\!\!\top}\widetilde{D}_{tzu}^{\top}\mathbf{Q_t}\mathbf{\overline{z}_t}\!-\!\mathbf{C_t}^{\!\!\!\top}\widetilde{D}_{tzc}^{\top}\mathbf{Q_t}\mathbf{\overline{z}_t} \nonumber\\
&\quad\!-\!\mathbf{\overline{z}_t}^{\!\!\top}\mathbf{Q_t}\widetilde{C}_{tz}x_{t}(0)\!-\!\mathbf{\overline{z}_t}^{\!\!\top}\mathbf{Q_t}\widetilde{D}_{tzu}\mathbf{u_t}\!-\!\mathbf{\overline{z}_t}^{\!\!\top}\mathbf{Q_t}\widetilde{D}_{tzc}\mathbf{C_{t}}\\
&\quad\!+\!\mathbf{\overline{z}_t}^{\top}\mathbf{Q_t}\mathbf{\overline{z}_t}. \nonumber\\
\end{aligned}
\end{equation*}

To satisfy the requirement of the convex optimisation, a Schur complement is used to rewrite the quadratic term $\mathbf{u_t}^{\!\!\!\top}\!\widetilde{D}_{tzu}^{\top}\!\!\mathbf{Q_t}\widetilde{D}_{tzu}\mathbf{u_t}$ in the 2,2-entity of matrix $L(\mathbf{u_t}, D_t,\overline{\gamma}_t)$ into a linear function in $\mathbf{u_t}$ as below:
\begin{equation*}
\begin{aligned}
    &L(\mathbf{u_t},D_t,\overline{\gamma}_t)
    =\\
    &\begin{bmatrix}
    -\widetilde{D}_{tzw}^{\top}\mathbf{Q_t}\widetilde{D}_{tzw}+
    D_t \!&\! -D_t\frac{\underline{\mathbf{w}}+\overline{\mathbf{w}}}{2}\!-\!bd_t
    \!&\! 0\\
    * \!&\! \underline{\mathbf{w}}^{\top}D_t\overline{\mathbf{w}}\!-\!cd_t+\overline{\gamma}_t&\mathbf{u_t}^{\top}\widetilde{D}_{tzu}^{\top}\mathbf{Q_t}\\
    *\!&\!*\!&\!\mathbf{Q_t}\end{bmatrix},
\end{aligned}
\end{equation*}
where $I \in \mathbb{R}^ {((N+1)\times 3) \times ((N+1)\times 3)}$ is an identity matrix. From \eqref{eq:RMPC_gamma_LHS}, it can be found that \eqref{eq:RMPC_gamma} is satisfied if and only if the following linear matrix inequality (LMI) is achieved,
\begin{equation}
    \begin{aligned}
    L(\mathbf{u_t},D_t,\overline{\gamma}_t)\succeq 0. \label{eq:RMPC_gamma_LMI1_time}
    \end{aligned}
\end{equation}

A similar SDPR technique is also applied to the stacked inequality constraints \eqref{eq:mpc_time_stack_constraint} to construct the corresponding LMIs. The LMI $L_{u}(\mathbf{u_t},D_{u_t,i})$ for the upper bound of \eqref{eq:mpc_time_stack_constraint} can be found through a subtraction between $\mathbf{f_t}$ and $\mathbf{\overline{f}_t}$,
\begin{equation}
\begin{aligned}
\label{eq:Robust_time_constraint_upper2}
    &e_{i}^{\top} \mathbf{f_t}-e_{i}^{\top} \mathbf{\overline{f}_t}
    \!=\!\\
    &\hspace{4mm}\!-\!(\mathbf{w}-\underline{\mathbf{w}})^{\top} D_{u_t,i}(\overline{\mathbf{w}}-\mathbf{w})\!-\!\left[\mathbf{w}^{\top} \,\,1\right]L_{u}(\mathbf{u_t},D_{u_t,i})\begin{bmatrix}\mathbf{w}\\1\end{bmatrix}\!\!\!,
\end{aligned}
\end{equation}
where $L_{u}(\mathbf{u_t},D_{u_t,i})=$
\begin{equation}
     \begin{bmatrix}
    D_{u_t,i}&-D_{u_t,i}\frac{\underline{\mathbf{w}}+\overline{\mathbf{w}}}{2}-\frac{\widetilde{D}_{tfw}^{\top}e_{i}}{2}\\
    * & \hspace{-2mm}\underline{\mathbf{w}}^{\top}D_{u_t,i}\overline{\mathbf{w}}\!-\!e_{i}^{\top}\!\left(\widetilde{C}_{tf}x_{t}(0)\!+\!\widetilde{D}_{tfu}\mathbf{u_t}\!+\!\widetilde{D}_{tfc}\mathbf{C_t}\!-\!\mathbf{\overline{f}_t}\right)
    \end{bmatrix}\!\!\!,   
\end{equation}
where $\widetilde{C}_{tf}$, $\widetilde{D}_{tfu}$, $\widetilde{D}_{tfc}$, $\widetilde{D}_{tfw}$ are stacked coefficient matrices after iterating the $f_t(k)$ equation in \eqref{eq:time_constraint_function_set}, 
$e_{i}$ is a unit vector whose $i$th element equals 1 and the rest of the elements are assigned to zero, and $i \in \{1, 2, \dots, N+1\}$, 
$D_{u_t,i}$ is a positive semi-definite diagonal matrix ($0\!\preceq\!D_{u_t,i} \in \mathbb{R}^ {N \times N}$). Since $-(\mathbf{w}-\underline{\mathbf{w}})^{\top} D_{u_t,i}(\overline{\mathbf{w}}-\mathbf{w})$ is non-positive, $LHS_{ \eqref{eq:Robust_time_constraint_upper2}}\leq 0$ is satisfied to the upper bound constraint in \eqref{eq:mpc_time_stack_constraint} if the matrix $L_{u}(\mathbf{u_t},D_{u_t,i})$ is positive semi-definite:
\begin{equation}
    L_{u}(\mathbf{u_t},D_{u_t,i})\succeq 0. \label{eq:RMPC_constraint_LMI_time1}
\end{equation}

By following the same deduction procedure from \eqref{eq:Robust_time_constraint_upper2}  to \eqref{eq:RMPC_constraint_LMI_time1}, the LMI $L_{l}(\mathbf{u_t},D_{u_t,i})$ for the lower bound of \eqref{eq:mpc_time_stack_constraint} can be found through a subtraction between $\mathbf{f_t}$ and $\mathbf{\underline{f}_t}$,
\begin{equation}
\begin{aligned}
\label{eq:Robust_time_constraint_lower2}
    &e_{i}^{\top} \mathbf{f_t}-e_{i}^{\top} \mathbf{\underline{f}_t}
    \!=\!\\
    &\hspace{5mm}-\!(\mathbf{w}-\underline{\mathbf{w}})^{\top} D_{l_t,i}(\overline{\mathbf{w}}-\mathbf{w})\!-\!\left[\mathbf{w}^{\top}\,\,1\right]L_{l}(\mathbf{u_t},D_{l_t,i})\begin{bmatrix}\mathbf{w}\\1\end{bmatrix}\!\!\!,\\
\end{aligned}
\end{equation}\\[-2.2ex]
and $L_{l}(\mathbf{u_t},D_{l_t,i})=$
\begin{equation}
\begin{bmatrix}
    D_{l_t,i}&-D_{l_t,i}\frac{\underline{\mathbf{w}}+\overline{\mathbf{w}}}{2}-\frac{\widetilde{D}_{tfw}^{\top}e_{i}}{2}\\
    * & \hspace{-2mm}\underline{\mathbf{w}}^{\top}D_{l,i}\overline{\mathbf{w}}\!-\!e_{i}^{\top}\left(\widetilde{C}_{tf}x_{t}(0)\!+\!\widetilde{D}_{tfu}\mathbf{u_t}\!+\!\widetilde{D}_{tfc}\mathbf{C_t}\!-\!\mathbf{\underline{f}_t}\right)
    \end{bmatrix}\!\!\!,
\end{equation}
where $D_{l_t,i}$ is a negative semi-definite diagonal matrix ($0\succeq D_{l_t,i} \in \mathbb{R}^ {N \times N}$).
Since $-(\mathbf{w}-\underline{\mathbf{w}})^{\top} D_{l_t,i}(\overline{\mathbf{w}}-\mathbf{w})$ is non-negative, $LHS_{\eqref{eq:Robust_time_constraint_lower2}}\geq 0$ is satisfied to the lower bound constraint in \eqref{eq:mpc_time_stack_constraint} if the matrix $L_{l}(\mathbf{u_t},D_{l_t,i})$ is negative semi-definite:
\begin{equation}
   L_{l}(\mathbf{u_t},D_{l_t,i})\preceq 0. \label{eq:RMPC_constraint_LMI_time2}
\end{equation}

To summarise, the RMPC of the convex optimal control problem in the time-domain \eqref{eq:mpc_time_stack_problem} can be achieved by solving the following convex optimisation problem:
\begin{subequations}\label{eq:RobustMPC_final_formula_time}
\begin{align}
     \min\limits_{\mathbf{u_t}}\quad & \overline{\gamma}_t\\
    \textbf{s.t. } & L(\mathbf{u_t},D,\overline{\gamma}_t)\!\!\succeq\!\! 0 ~\eqref{eq:RMPC_gamma_LMI1_time},\,L_u(\mathbf{u_t},D_{u_t,i})\!\!\succeq\!\! 0 ~\eqref{eq:RMPC_constraint_LMI_time1}, \nonumber\\&L_l(\mathbf{u_t},D_{l_t,i})\!\!\preceq\!\! 0 ~\eqref{eq:RMPC_constraint_LMI_time2}, \\
    \textbf{given: }& x_t(0)\!=\![v(0),\,\Delta s (0)]^{\top},\\
    & 0\preceq D_t \in \mathbb{R}^ {N \times N}, \,0\preceq D_{u_t,i}\in \mathbb{R}^ {N \times N},\,\nonumber\\
    & 0\succeq D_{l_t,i}\in \mathbb{R}^ {N \times N}. 
\end{align}
\end{subequations}

\section{Simulation Results}
\label{sec:simulation results}
The performance of the proposed RMPC on the car-following scenario is evaluated in twofold in this section, 1) an investigation of the validity of the proposed RMPC scheme in the time-domain by comparing it with a nominal MPC under the same conditions,
2) a comparison of the performance between the proposed RMPC scheme in the time-domain and the space-domain formulated benchmark RMPC scheme presented in~\cite{9838441} under the same initial conditions. 
All the convex optimisation problems are solved by the Yalmip toolkit with MOSEK solver in the Matlab environment on a 1.6 GHz Dual-Core Intel Core i5 processor with 8 GB memory. The sampling interval of the proposed time-domain RMPC is kept the same for all cases at $\delta t\!=\!0.2$~s.
Before the analysis of numerical examples, the initial setup for the simulation in this work is first introduced.
\subsection{Simulation Setup}
Although the nominal model of the ego vehicle is considered in \eqref{eq:nonlinear_v}, this work specifies the bounds of disturbance $w(k)$ by entailing
the varied road slope and coefficients of rolling and air-drag resistances in practice. The subtraction of the vehicle model \eqref{eq:nonlinear_v} using nominal and varied resistance parameters derives a mathematics description of $w(k)$, as follows:\\[-2.2ex]
\begin{equation}
\label{eq:disturbance_formula}
\begin{aligned}
w(k)=&\frac{\tilde{f}_dv(k)^2}{m}+g\tilde{f}_r  \\
&-\frac{f_d(k)v(k)^2}{m}-gf_r(k)\cos\left(\theta(k)\right)-g\sin\left(\theta(k)\right),  \\ 
=&\left(\tilde{f}_d\!-\!f_d(k)\right)\frac{v(k)^2}{m}\!+\!g\tilde{f}_r\!-\!G(k)\cos\left(\theta(k)\!-\!\phi(k)\right),
\end{aligned}
\end{equation}\\[-2.2ex]
where $G(k)=g\sqrt{f_r(k)^2+1}$, and $\phi(k)=\arctan(\frac{1}{f_r(k)})$.
From \eqref{eq:disturbance_formula}, the bounds $\underline{w}$ and $\overline{w}$ of $w(k)$ can be determined through a conservative relaxation with consideration of the worst-case scenario, leading to:
\begin{subequations}\label{eq:disturbance_boundary_wortcases}
\begin{align}
\overline{w}&=\left(\tilde{f}_d-\underline{f}_d\right)\frac{v_{\max}^2}{m}+g\tilde{f}_r-g\underline{f}_{r}\cos(\underline{\theta})-g\sin(\underline{\theta}) ,\label{eq:disturbance_upper_bar} 
\\
\underline{w}&=\left(\tilde{f}_d-\overline{f}_{d} \right)\frac{v_{\max}^2}{m}+g\tilde{f}_r-g\overline{f}_{r}\cos(\overline{\theta})-g\sin(\overline{\theta}), \label{eq:disturbance_lower_bar}
\end{align}
\end{subequations}
where the values of the bounds of the air-drag resistance coefficient, $\underline{f}_{d}$ and $\overline{f}_{d}$, are determined based on the analysis of the coefficient of air-drag resistance with respect to the headway distance in~\cite{8796226}. The values of the bounds of the rolling resistance coefficient, $\underline{f}_{r}$ and $\overline{f}_{r}$, are determined based on the investigation using a nominal rolling coefficient~\cite{doi:10.1080/14680629.2016.1160835}, and the values of the bounds $\underline{\theta}$ and $\overline{\theta}$ of the road slope are determined through
the consideration to bear a certain degree of slope in practice~\cite{LIU201774}.


The battery energy consumption produced by the solution of the proposed method and the benchmark solution is evaluated using a unique, and widely used motor efficiency map for a fair comparison between the two approaches.
This post-evaluation of the battery energy consumption entails
energy losses during processes of energy transmission and regeneration.
The actual battery energy consumption of the ego vehicle over the time window $k\!\in\![0,\bar{k}]$ follows:
\begin{equation}\label{eq:Jb_evaluation}
\begin{aligned}
E^*_{Bat}\!=\!\sum_{k=0}^{\bar{k}}P_{\text{Bat},i}^*(k)\,\delta t,
 \end{aligned}
\end{equation}
with
\[
P_{\text{Bat}}^*(k)\!=\! \left\{
 \begin{array}{ll}
   \displaystyle \!\!\!\frac{F_{w}^*(k)}{\eta_m(F_{w}^*(k),v^*(k))}, & \forall F_{w}(k)\geq0, \\
   \displaystyle
    \!\!\!{F_{w}^*(k)}{\eta_m(F_{w}^*(k),v^*(k))}, & \forall F_{w}(k)<0,
 \end{array}\right.
\]
where $F_{w}^*(k)$ is the optimal input force on wheels, $v^*(k)$ is the optimal speed, and $\eta_m(F_{w}^*(k),v^*(k))$ is the powertrain efficiency provided by a look-up table from Advisor~\cite{MARKEL2002255}.

\begin{figure}[t!]
\centering
\includegraphics[width=\columnwidth]{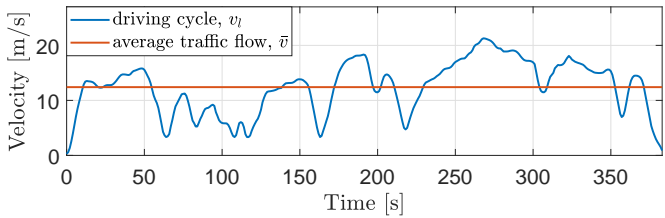}\\[-1ex]
\caption{Driving cycle of WLTP-M Phase.}
\label{fig:driving_cycle}
\end{figure}

The speed profile of the leading vehicle $v_l(k)$ follows the medium phase of the worldwide harmonised light vehicles test cycles (WLTP-M) to emulate urban driving, with its average speed set as the cruise speed $\bar{v}$ (see Fig.~\ref{fig:driving_cycle}).
The main characteristic parameters of the models of the electric vehicle and the regulations of the car-following paradigm for WLTP-M driving cycle are summarised in Table.~\ref{tab:model_specification_parameters}.

\begin{table}[t!]
    \centering
        \caption{\textsc{Parameters of car-following model.}}
    \label{tab:model_specification_parameters}
    \begin{tabular*}{1\columnwidth}{l @{\extracolsep{\fill}} c@{\extracolsep{\fill}}c}
        \hline
        \hline
          Description & Symbols & Values \\
         \hline
         Earth gravity& $g$ & $9.8~m/s^2$ \\
         
         Standstill distance & $s_{0}$ & $2~m$ \\
         
         Nominal rolling resistance coefficient& $\tilde{f}_{r}$ & $0.01$ \\
         Nominal air-drag resistance coefficient& $\tilde{f}_{d}$ & $0.34$ \\

         Ego vehicle initial velocity& $v_0$ & $0.2778~m/s$\\
         
         Minimum/maximum velocity limits & $v_{\min/\max}$ & $0$/$22.352~m/s$   \\ 
         
         Minimum/maximum force on wheels & $F_{w,\min/\max}$&-7800/3500 N    \\ 
         
         
         Minimum/maximum time difference& $\Delta t_{\min}$/$\Delta t_{\max}$ & $1$/$8~s$\\
         
         Ego vehicle mass& $m$ & $1200~kg$\\
         
         Minimum acceleration of the ego vehicle  & $a_{\min}$ & -6.5 $m/s^2$  \\ 
         
         Initial distance gap  & $\Delta s(0)$ & $3~m$\\
         
         
         Bounds of air-drag resistance coefficient & $\underline{f}_d$/$\overline{f}_d$ &0.296/0.380 \\
         
         Bounds of rolling resistance coefficient & $\underline{f}_r$/$\overline{f}_r$ & 0.008/0.012 \\
         
         Bounds of road slope & $\underline{\theta}$/$\overline{\theta}$ & $-0.573^{\circ}$/$0.573^{\circ}$\\  
        
         Bounds of disturbance & $\underline{w}$/$\overline{w}$ & -0.134/0.136\\ 

         \hline
         \hline
    \end{tabular*}
\end{table}

\subsection{Numerical Examples}
\begin{figure}[t!]
\centering
\includegraphics[width=\columnwidth]{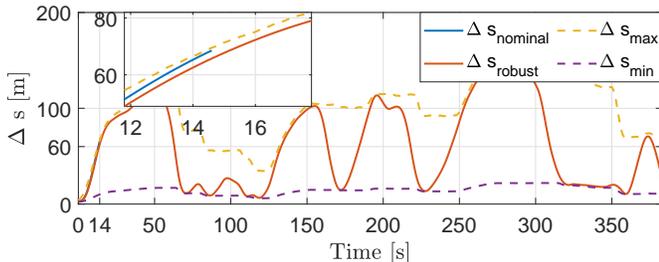}\\[-1.6ex]
\caption{Comparison on the inter-vehicular distance between nominal and robust MPCs}
\label{fig:nominal_gap_compare}
\end{figure}

The robustness and the validity of the proposed method are first investigated against the results derived from a nominal MPC under the same conditions and disturbances. As shown in Fig.~\ref{fig:nominal_gap_compare}, with the additive disturbance, an infeasible solution is found in the trajectory of the inter-vehicular distance of the nominal MPC case (represented by a blue curve), which ends up at around $14$~s due to the violation of the upper bound constraint. 
However, the proposed RMPC is able to maintain a gap to the constraint borders such that optimal feasible control sequences can be derived even with the additive disturbance. Moreover, the optimal results of the inter-vehicular distance for the RMPC method tend to reach the maximum and minimum allowed bounds. This could encourage the ego vehicle to optimally adjust its speed through full utilisation of the constrained headway distance range for the energy-saving purpose in~\eqref{eq:time_cost_function_set}.
Overall, the robustness and validity of the proposed method are demonstrated.

\begin{figure}[t!]
\centering
\includegraphics[width=\columnwidth]{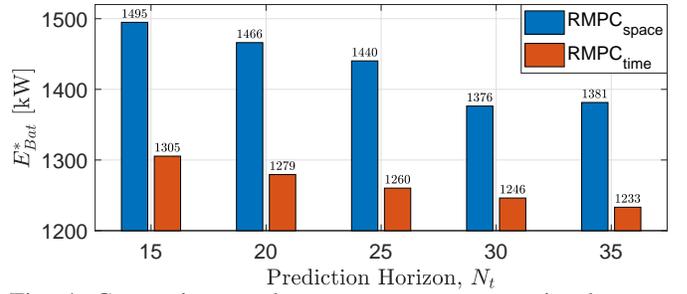}\\[-1.6ex]
\caption{Comparison on battery energy consumption between robust time-domain and space-domain schemes.}
\label{fig:robust_energy_compare}
\end{figure}

The comparison of the battery energy consumption between the proposed time-domain formulated method and the space-domain formulated benchmark for different prediction horizon lengths from $N\!=\!15$ to $N\!=\!35$ is presented in Fig.~\ref{fig:robust_energy_compare}. As can be seen, the time-domain scheme can save more energy than the space-domain scheme under the same disturbance influence for all prediction horizon lengths chosen in this example case. The average energy consumption improvement of the proposed method against the benchmark is 11.6\%. Although the benchmark utilises a direct energy consumption model in its objective function, its optimal control sequence exists larger acceleration and deceleration rates as compared to the results of the proposed method minimising the L$^2$-norm of input force. As a result, this manoeuvre behaviour of the benchmark leads to an increase in energy consumption. The performance of the acceleration and deceleration rate of the optimal results derived by two methods is investigated through an analysis of the root mean square (RMS) of the jerk variable, which is the derivative of the optimal acceleration and deceleration in time, $j\!=\!\frac{d^2v}{dt^2}$,
\begin{equation}
\label{eq:jerk_RMS}
    j_{RMS}=\sqrt{\frac{1}{\bar{k}}\sum_{i=1}^{\bar{k}}j(i)^2}.
\end{equation}\\[-1.5ex]
The comparison of the RMS-jerk between the two methods is shown in Table~\ref{tab:robust_jerk_compare}, where $j_{\text{RMS}}$ of the time-domain scheme is found to be always smaller than that of the space-domain scheme. This founding verifies the fact that harsh acceleration and deceleration rates are found in the space-domain formulated benchmark, which results in extra energy consumption. Besides, the smaller $j_{\text{RMS}}$ of the time-domain scheme also indicates that the proposed RMPC in this work has a higher passenger travel comfort as compared to the benchmark as harsh acceleration and deceleration are avoided.

\begin{table}[!t]
\centering 
\caption{\textsc{ Comparison on jerk between robust time-domain and space-domain schemes.}}
    \label{tab:robust_jerk_compare}
\begin{tabular*}{1\columnwidth}{l @{\extracolsep{\fill}}  c @{\extracolsep{\fill}}c @{\extracolsep{\fill}} c @{\extracolsep{\fill}} c @{\extracolsep{\fill}}   c}
\hline
\hline
Prediction horizon length, $N$ & 15   & 20   & 25   & 30  & 35 \\
 \hline
$j_{\text{RMS}}$ by the time-domain scheme& 0.573 &0.562&0.557&0.555&0.553\\
\hline
$j_{\text{RMS}}$ by the space-domain scheme& 2.257&2.157&2.001&1.334&1.481\\
\hline
\hline
\end{tabular*}
\end{table}

\section{Conclusion and Future Work}\label{sec:conclusions}
The paper addresses a nonlinear car-following problem under vehicular ad-hoc communication (VANET) framework by proposing a time-domain formulated convex RMPC scheme. The nonlinearity existing in the car-following problem is convexified through feedback linearisation of the nonlinear dynamics, such as the air drag resistance. The performance of the proposed scheme is compared with a space-domain formulated benchmark in the authors' previous work. Numerical examples validate the effectiveness of the proposed convex RMPC with additive disturbances. The comparison between the proposed scheme and benchmark shows that the ego vehicle can achieve higher energy efficiency and passenger comfort level in the proposed time-domain formulated RMPC framework. Future research focuses on two aspects, 1) a consideration of both energy losses and battery thermal effect in the system formulation to emulate a more practical driving situation, 2) an expansion of uncertainty types, including both modelling mismatches of the ego vehicle and the communication error and delay of the leading vehicle information.





\bibliographystyle{IEEEtran}
\bibliography{reference}

\begin{thebibliography}{10}
\providecommand{\url}[1]{#1}
\csname url@samestyle\endcsname
\providecommand{\newblock}{\relax}
\providecommand{\bibinfo}[2]{#2}
\providecommand{\BIBentrySTDinterwordspacing}{\spaceskip=0pt\relax}
\providecommand{\BIBentryALTinterwordstretchfactor}{4}
\providecommand{\BIBentryALTinterwordspacing}{\spaceskip=\fontdimen2\font plus
\BIBentryALTinterwordstretchfactor\fontdimen3\font minus
  \fontdimen4\font\relax}
\providecommand{\BIBforeignlanguage}[2]{{%
\expandafter\ifx\csname l@#1\endcsname\relax
\typeout{** WARNING: IEEEtran.bst: No hyphenation pattern has been}%
\typeout{** loaded for the language `#1'. Using the pattern for}%
\typeout{** the default language instead.}%
\else
\language=\csname l@#1\endcsname
\fi
#2}}
\providecommand{\BIBdecl}{\relax}
\BIBdecl

\bibitem{guanetti2018control}
J.~Guanetti, Y.~Kim, and F.~Borrelli, ``Control of connected and automated
  vehicles: State of the art and future challenges,'' \emph{Annual reviews in
  control}, vol.~45, pp. 18--40, 2018.

\bibitem{huang2018ecological}
K.~Huang, X.~Yang, Y.~Lu, C.~C. Mi, and P.~Kondlapudi, ``Ecological driving
  system for connected/automated vehicles using a two-stage control
  hierarchy,'' \emph{IEEE Transactions on Intelligent Transportation Systems},
  vol.~19, no.~7, pp. 2373--2384, 2018.

\bibitem{bae2019design}
S.~Bae, Y.~Kim, J.~Guanetti, F.~Borrelli, and S.~Moura, ``Design and
  implementation of ecological adaptive cruise control for autonomous driving
  with communication to traffic lights,'' in \emph{2019 American Control
  Conference (ACC)}.\hskip 1em plus 0.5em minus 0.4em\relax IEEE, 2019, pp.
  4628--4634.

\bibitem{xu2021comparison}
H.~Xu, C.~G. Cassandras, L.~Li, and Y.~Zhang, ``Comparison of cooperative
  driving strategies for {CAVs} at signal-free intersections,'' \emph{IEEE
  Transactions on Intelligent Transportation Systems}, 2021.

\bibitem{liu2020high}
C.~Liu, Y.~Zhang, T.~Zhang, X.~Wu, L.~Gao, and Q.~Zhang, ``High throughput
  vehicle coordination strategies at road intersections,'' \emph{IEEE
  Transactions on Vehicular Technology}, vol.~69, no.~12, pp. 14\,341--14\,354,
  2020.

\bibitem{jia2021enhanced}
D.~Jia, H.~Chen, Z.~Zheng, D.~Watling, R.~Connors, J.~Gao, and Y.~Li, ``An
  enhanced predictive cruise control system design with data-driven traffic
  prediction,'' \emph{IEEE Transactions on Intelligent Transportation Systems},
  2021.

\bibitem{7225178}
M.~Vajedi and N.~L. Azad, ``Ecological adaptive cruise controller for plug-in
  hybrid electric vehicles using nonlinear model predictive control,''
  \emph{IEEE Transactions on Intelligent Transportation Systems}, vol.~17,
  no.~1, pp. 113--122, 2016.

\bibitem{GUO2022124732}
C.~Guo, C.~Fu, R.~Luo, and G.~Yang, ``Energy-oriented car-following control for
  a front- and rear-independent-drive electric vehicle platoon,''
  \emph{Energy}, vol. 257, p. 124732, 2022.

\bibitem{sun2022tube}
H.~Sun, L.~Dai, and B.~Chen, ``Tube-based distributed model predictive control
  for heterogeneous vehicle platoons via convex optimization,'' in \emph{2022
  IEEE 25th International Conference on Intelligent Transportation Systems
  (ITSC)}.\hskip 1em plus 0.5em minus 0.4em\relax IEEE, 2022, pp. 1122--1127.

\bibitem{8957499}
J.~Hu, P.~Bhowmick, F.~Arvin, A.~Lanzon, and B.~Lennox, ``Cooperative control
  of heterogeneous connected vehicle platoons: An adaptive leader-following
  approach,'' \emph{IEEE Robotics and Automation Letters}, vol.~5, no.~2, pp.
  977--984, 2020.

\bibitem{9838441}
S.~Yu, X.~Pan, A.~Georgiou, B.~Chen, I.~M. Jaimoukha, and S.~A. Evangelou,
  ``Robust model predictive control framework for energy-optimal adaptive
  cruise control of battery electric vehicles,'' in \emph{2022 European Control
  Conference (ECC)}, 2022, pp. 1728--1733.

\bibitem{Pan_2022}
X.~Pan, B.~Chen, S.~Timotheou, and S.~A. Evangelou, ``A convex optimal control
  framework for autonomous vehicle intersection crossing,'' \emph{{IEEE}
  Transactions on Intelligent Transportation Systems}, 2022.

\bibitem{7659496}
A.~Loulizi, Y.~Bichiou, and H.~Rakha, ``Steady-state car-following time gaps:
  An empirical study using naturalistic driving data,'' \emph{Journal of
  Advanced Transportation}, vol. 2019, pp. 1--9, 2019.

\bibitem{7374229}
C.~B. Math, A.~Ozgur, S.~H. de~Groot, and H.~Li, ``Data rate based congestion
  control in {V2V} communication for traffic safety applications,'' in
  \emph{2015 IEEE Symposium on Communications and Vehicular Technology in the
  Benelux (SCVT)}, 2015, pp. 1--6.

\bibitem{pan2022TCST}
X.~Pan, B.~Chen, L.~Dai, S.~Timotheou, and S.~A. Evangelou, ``A hierarchical
  robust control strategy for decentralized signal-free intersection
  management,'' \emph{arXiv preprint arXiv:2206.14986}, 2022.

\bibitem{9151336}
N.~K. Sharma, A.~Hamednia, N.~Murgovski, E.~R. Gelso, and J.~Sjöberg,
  ``Optimal eco-driving of a heavy-duty vehicle behind a leading heavy-duty
  vehicle,'' \emph{IEEE Transactions on Intelligent Transportation Systems},
  vol.~22, no.~12, pp. 7792--7803, 2021.

\bibitem{9631299}
Y.~Lee, D.~Y. Lee, S.~H. Lee, and Y.~Kim, ``A comparative study on model
  predictive control design for highway car-following scenarios: Space-domain
  and time-domain model,'' \emph{IEEE Access}, vol.~9, pp. 162\,291--162\,305,
  2021.

\bibitem{rawlings2017model}
J.~Rawlings, D.~Mayne, and M.~Diehl, \emph{Model Predictive Control: Theory,
  Computation, and Design}.\hskip 1em plus 0.5em minus 0.4em\relax Nob Hill
  Publishing, 2017.

\bibitem{Tahir13}
F.~Tahir and I.~{M. Jaimoukha}, ``Causal state-feedback parameterizations in
  robust model predictive control,'' \emph{Automatica}, vol.~49, pp.
  2675--2682, 2013.

\bibitem{9304474}
A.~Georgiou, F.~Tahir, S.~A. Evangelou, and I.~M. Jaimoukha, ``Robust moving
  horizon state estimation for uncertain linear systems using linear matrix
  inequalities,'' in \emph{2020 59th IEEE Conference on Decision and Control
  (CDC)}, 2020, pp. 2900--2905.

\bibitem{GEORGIOU202011974}
A.~Georgiou, S.~A. Evangelou, I.~M. Jaimoukha, and G.~Downton, ``Tracking
  control for directional drilling systems using robust feedback model
  predictive control,'' \emph{IFAC-PapersOnLine}, vol.~53, no.~2, pp.
  11\,974--11\,981, 2020.

\bibitem{8796226}
D.~R. Lopes and S.~A. Evangelou, ``Energy savings from an eco-cooperative
  adaptive cruise control: a {BEV} platoon investigation,'' in \emph{2019 18th
  European Control Conference (ECC)}, 2019, pp. 4160--4167.

\bibitem{doi:10.1080/14680629.2016.1160835}
J.~A. Ejsmont, G.~Ronowski, B.~Świeczko Żurek, and S.~Sommer, ``Road texture
  influence on tyre rolling resistance,'' \emph{Road Materials and Pavement
  Design}, vol.~18, no.~1, pp. 181--198, 2017.

\bibitem{LIU201774}
K.~Liu, T.~Yamamoto, and T.~Morikawa, ``Impact of road gradient on energy
  consumption of electric vehicles,'' \emph{Transportation Research Part D:
  Transport and Environment}, vol.~54, pp. 74--81, 2017.

\bibitem{MARKEL2002255}
T.~Markel, A.~Brooker, T.~Hendricks, V.~Johnson, K.~Kelly, B.~Kramer,
  M.~O’Keefe, S.~Sprik, and K.~Wipke, ``Advisor: a systems analysis tool for
  advanced vehicle modeling,'' \emph{Journal of Power Sources}, vol. 110,
  no.~2, pp. 255--266, 2002.

\end{thebibliography}

\end{document}